\documentclass[twocolumn,secnumarabic,amssymb, nobibnotes, aps, prd]{revtex4-2}
\usepackage[pdftex]{graphicx}
\usepackage{braket}

\begin{document}

\title{Generation of a precise time scale assisted by a near-continuously operating optical lattice clock}%

\author{Takumi Kobayashi$^{1}$}%
\thanks{These two authors contributed equally to this work. Email: takumi-kobayashi@aist.go.jp}
\author{Daisuke Akamatsu$^{2}$}
\thanks{These two authors contributed equally to this work. Email: takumi-kobayashi@aist.go.jp}
\author{Kazumoto Hosaka$^{1}$}
\author{Yusuke Hisai$^{2\dag}$}
\author{Akiko Nishiyama$^{1}$}
\author{Akio Kawasaki$^{1}$}
\author{Masato Wada$^{1}$}
\author{Hajime Inaba$^{1}$}
\author{Takehiko Tanabe$^{1}$}
\author{Feng-Lei Hong$^{2}$}
\author{Masami Yasuda$^{1}$}
\affiliation{$^{1}$National Metrology Institute of Japan (NMIJ), National Institute of Advanced Industrial Science and Technology (AIST), 1-1-1 Umezono, Tsukuba, Ibaraki 305-8563, Japan \\ $^{2}$Department of Physics, Graduate School of Engineering Science, Yokohama National University, 79-5 Tokiwadai, Hodogaya-ku, Yokohama 240-8501, Japan\\$^{\dag}$Present address: Shimadzu Corporation, 3-9-4 Hikaridai, Seika-cho, Soraku-gun, Kyoto 619-0237, Japan}
\begin{abstract}  
We report on a reduced time variation of a time scale with respect to Coordinated Universal Time (UTC) by steering a hydrogen-maser-based time scale with a near-continuously operating optical lattice clock. The time scale is generated in a post-processing analysis for 230 days with a hydrogen maser with its fractional frequency stability limited by a flicker floor of $2\times10^{-15}$ and an Yb optical lattice clock operated with an uptime of 81.6 $\%$. During the 230-day period, the root mean square time variation of our time scale with respect to UTC is 0.52 ns, which is a better performance compared with those of time scales steered by microwave fountain clocks that exhibit root mean square variations from 0.99 ns to 1.6 ns. With the high uptime achieved by the Yb optical lattice clock, our simulation implies the potential of generating a state-of-the-art time scale with a time variation of $<0.1$ ns over a month using a better hydrogen maser reaching the mid $10^{-16}$ level. This work demonstrates that a use of an optical clock with a high uptime enhances the stability of a time scale.
\end{abstract}
\maketitle

\section{Introduction}
The generation of a time scale is an essential mission in national metrology institutes for time keeping and frequency calibration services for industries, and is also crucial for navigation, telecommunication, and new potential applications including geodesy \cite{Takano2016,Grotti2018} and studies on fundamental physics \cite{Hees2016,Wcilso2018,Ashby2018,Kennedy2020,Takamoto2020,Lange2021,Kobayashi2022,Filzinger2023,Sherrill2023}. A time scale is generated by a continuously running flywheel oscillator with its frequency steered by an atomic frequency standard. As a global time scale, Bureau International des Poids et Mesures (BIPM) computes International Atomic Time (TAI) and Coordinated Universal Time (UTC), which are generated by several hundreds of microwave flywheel clocks around the world and steered by Cs and Rb microwave clocks and Yb and Sr optical lattice clocks that are officially approved as primary and secondary frequency standards \cite{Panfilo2019,CircularT}. A national institute $k$ generates its official local time scale UTC($k$) by synchronizing a local flywheel to UTC using the time comparison data obtained via a satellite link. For example, National Metrology Institute of Japan (NMIJ) maintains UTC(NMIJ) within several ten nanoseconds from UTC in this way.

The time difference of UTC($k$) with respect to UTC is expected to be reduced by locally realizing a time scale based on a primary or secondary frequency standard instead of referencing only UTC, since the time comparison data between UTC and UTC($k$) are provided by BIPM with a latency of about one month. The improvements of UTC($k$) in this way have so far been demonstrated by steering flywheel hydrogen masers to Cs or Rb microwave fountain clocks \cite{Bauch2012,Domnin2013,Peil2014,Rovera2016}. 

Towards a redefinition of the SI second \cite{Riehle2018,Hong2016,Dimarcq2023}, some institutes have started to incorporate optical clocks in local time scales. The best optical clocks reach the fractional frequency stabilities and uncertainties at the $10^{-18}-10^{-19}$ levels \cite{Bloom2014,Ushijima2015,Huntemann2016,McGrew2018,Brewer2019,Hung2022,Zhang2022,Zhiqiang2023}, and thus they have potentials to significantly improve the stabilities of time scales compared with microwave clocks. Since the availabilities of optical clocks are commonly limited for short periods with low uptimes, previous works mostly rely on microwave \cite{Grebing2016,Hachisu2018,Yao2019,Formichella2021,Zhu2022,Lee2023,Xu2023} and optical \cite{Milner2019} flywheels with good frequency stabilities to bridge the gaps in the operations of optical clocks. Among the previous works, long-term generations of optical-clock-based time scales for several months are reported with microwave flywheels that reach the low $10^{-16}$ levels. One example is a time scale based on a hydrogen maser steered by a Sr optical lattice clock operated for $10^{4}$ s per week, which exhibits a time difference of 0.79 ns after 5 months from another stable global time scale TT(BIPM) \cite{Hachisu2018}. Another example is a time scale based on an ensemble of a few hydrogen masers and a few commercial Cs clocks steered by an Yb optical lattice clock with an uptime of 6 $\%$, which follows UTC with a root mean square time variation of 0.40 ns for 160 days \cite{Yao2019}. In the previous methods, however, the stabilities of time scales are ultimately limited by stochastic fluctuations of the flywheels during the dead time of the optical clocks. Therefore, the continuous operation of an optical clock is desirable to take full advantage of its stability. 

For potential applications of optical clocks including the TAI calibration and geodesy, several groups have developed robust optical clocks and reported their operations with high uptimes \cite{Hill2016,Lodewyck2016,Baynham2017,Stuhler2021,Zeng2023} (e.g., 84 $\%$ for 25 days \cite{Hill2016}). So far, their high uptime operation periods typically last for $\lesssim1$ month, which is not practically long enough in terms of the generation of a time scale. However, NMIJ has achieved the operation of an Yb optical lattice clock with an uptime of $>80$ $\%$ for many months \cite{Kobayashi2020}. Thus, it is worth discussing the method to generate a time scale based on the nearly continuous operation of an optical clock.

In this work, we report on post-processing generation of a local time scale UTC(NMIJ)$^{\prime}$ for 230 days by steering the frequency of a single hydrogen maser to that of an Yb optical lattice clock operated with an uptime of 81.6 $\%$ using a Kalman filter algorithm. The root mean square time variation of UTC(NMIJ)$^{\prime}$ with respect to UTC is 0.52 ns. Although the fractional frequency stability of our hydrogen maser is limited by a flicker floor of $2\times10^{-15}$, UTC(NMIJ)$^{\prime}$ exhibits a better performance compared with those of time scales based on Cs or Rb fountain clocks. We also simulate a projected performance of UTC(NMIJ)$^{\prime}$ with a better flywheel hydrogen maser that reaches $5\times10^{-16}$, which implies a time variation of $<0.1$ ns over a month.

\section{Method}
\subsection{Experimental setup}
\label{experimentalsetupsection}
\begin{figure}[t]
\includegraphics[scale=0.38,angle=0]{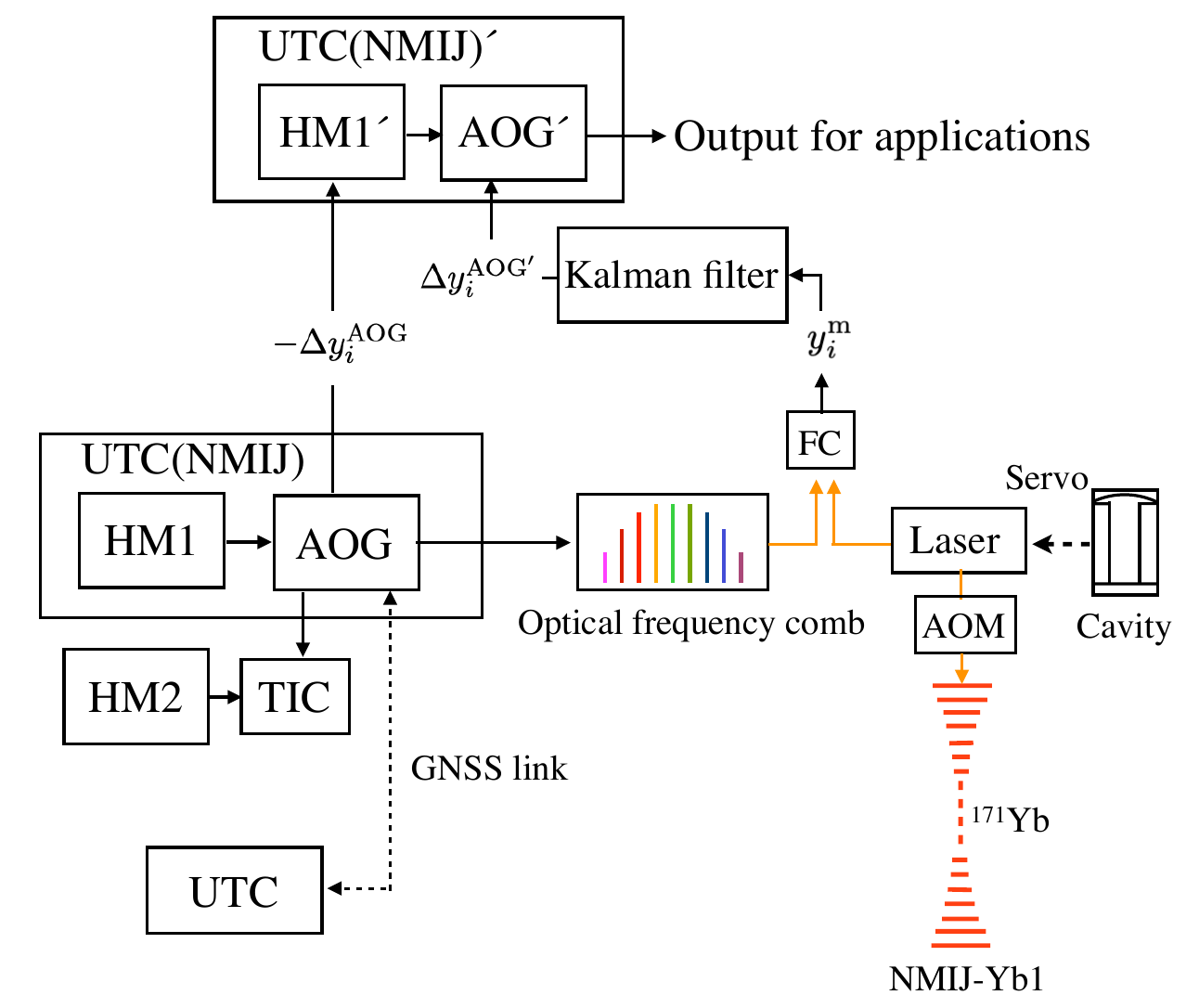}
\caption{Experimental setup for generating a local time scale UTC(NMIJ)$^{\prime}$ based on the Yb optical lattice clock NMIJ-Yb1. HM: hydrogen maser, AOG:  auxiliary output generator, AOM: acousto optic modulator, FC: frequency counter, TIC: time interval counter, GNSS: global navigation satellite systems, $y^{\mathrm{m}}_{i}$: fractional frequency difference between HM1$^{\prime}$ and NMIJ-Yb1, $\Delta y^{\mathrm{AOG}}$: fractional frequency correction applied to AOG, $\Delta y^{\mathrm{AOG^{\prime}}}$: fractional frequency correction applied to AOG$^{\prime}$.}
\label{experimentalsetup}
\end{figure}

Figure \ref{experimentalsetup} shows the experimental setup for generating UTC(NMIJ)$^{\prime}$. Our current official time scale UTC(NMIJ) is generated by a hydrogen maser HM1 (KVARZ, CH1-75A) and an auxiliary output generator AOG (SYMMETRICOM AOG-110) for frequency steering. The frequency of UTC(NMIJ) is steered typically once every few months by manually adding the fractional correction frequency $\Delta y^{\mathrm{AOG}}\sim10^{-15}$ to the frequency of AOG to reduce the time difference between UTC(NMIJ) and UTC. For post-processing generation of UTC(NMIJ)$^{\prime}$, we compute a virtual hydrogen maser HM1$^{\prime}$ by subtracting $\Delta y^{\mathrm{AOG}}$ from UTC(NMIJ) so that HM1$^{\prime}$ does not have frequency steps by the manual steering and thus has frequency stability mostly equivalent to that of HM1. UTC(NMIJ)$^{\prime}$ is generated by HM1$^{\prime}$ and another virtual AOG for steering UTC(NMIJ)$^{\prime}$, denoted by AOG$^{\prime}$. The contribution of AOG to the fractional instability of UTC(NMIJ) is confirmed to be $\sim1\times10^{-16}$ at an averaging time $\tau\sim2\times10^{6}$ s, limited by measurement noise of  $2\times10^{-10}/(\tau/\mathrm{s})$ of a time interval counter. We expect a lower contribution of AOG according to its catalog specification ($3\times10^{-13}/(\tau/\mathrm{s})$).

The frequency of HM1$^{\prime}$ is compared with that of the Yb optical lattice clock NMIJ-Yb1 \cite{Kobayashi2018,Kobayashi2020,Kobayashi2022} using an optical frequency comb \cite{Inaba2006}. The beat frequency between a laser stabilized to an ultra-stable optical cavity and the comb phase-locked to UTC(NMIJ) is measured by a frequency counter. The frequency of the laser is shifted by an acousto optic modulator (AOM) and stabilized to the $^{1}\mathrm{S}_{0}$-$^{3}\mathrm{P}_{0}$ clock transition of $^{171}$Yb. The beat and AOM frequencies are used to calculate the fractional frequency difference $y^{\mathrm{m}}_{i}$ between HM1$^{\prime}$ and NMIJ-Yb1 at a time epoch $i$ by the approximation
\begin{equation}
y^{\mathrm{m}}_{i}\sim \frac{f^{\mathrm{a}}_{\mathrm{UTC(NMIJ)}}/f^{\mathrm{a}}_{\mathrm{Yb}}}{f^{\mathrm{n}}_{\mathrm{UTC(NMIJ)}}/f^{\mathrm{n}}_{\mathrm{Yb}}}-1-\Delta y^{\mathrm{AOG}}_{i},
\label{hm1ybeq}
\end{equation}
where $f^{\mathrm{a(n)}}_{\mathrm{X}}$ denotes the actual (nominal) frequency of $\mathrm{X}$ and $\Delta y^{\mathrm{AOG}}_{i}$ the correction frequency applied to AOG at the epoch $i$. The nominal frequency is chosen as $f^{\mathrm{n}}_{\mathrm{UTC(NMIJ)}}=10$ MHz and $f^{\mathrm{n}}_{\mathrm{Yb}}=518\,295\,836\,590\,863.6$ Hz which is the CIPM (Comit\'{e} International des Poids et Mesures) recommended frequency during the period of the generation of UTC(NMIJ)$^{\prime}$ \cite{Riehle2018}. The systematic frequency shift of NMIJ-Yb1 is evaluated with a fractional uncertainty of $4.0\times10^{-16}$ \cite{Kobayashi2020}, and is corrected in the calculation of $y^{\mathrm{m}}_{i}$. The frequency stability of NMIJ-Yb1 is evaluated as $7.1\times10^{-15}/\sqrt{(\tau/\mathrm{s})}$ by comparing it with a Sr optical lattice clock \cite{Akamatsu2014,Hisai2021,stabilitynote}. The uncertainty of the comb-based frequency measurement is estimated to be $2.2\times10^{-16}$ from a flicker noise observed when comparing two independent combs at $\tau\sim(1-5)\times10^{4}$ s \cite{Kobayashi2020}. 

\begin{figure}[t]
\includegraphics[scale=0.38,angle=0]{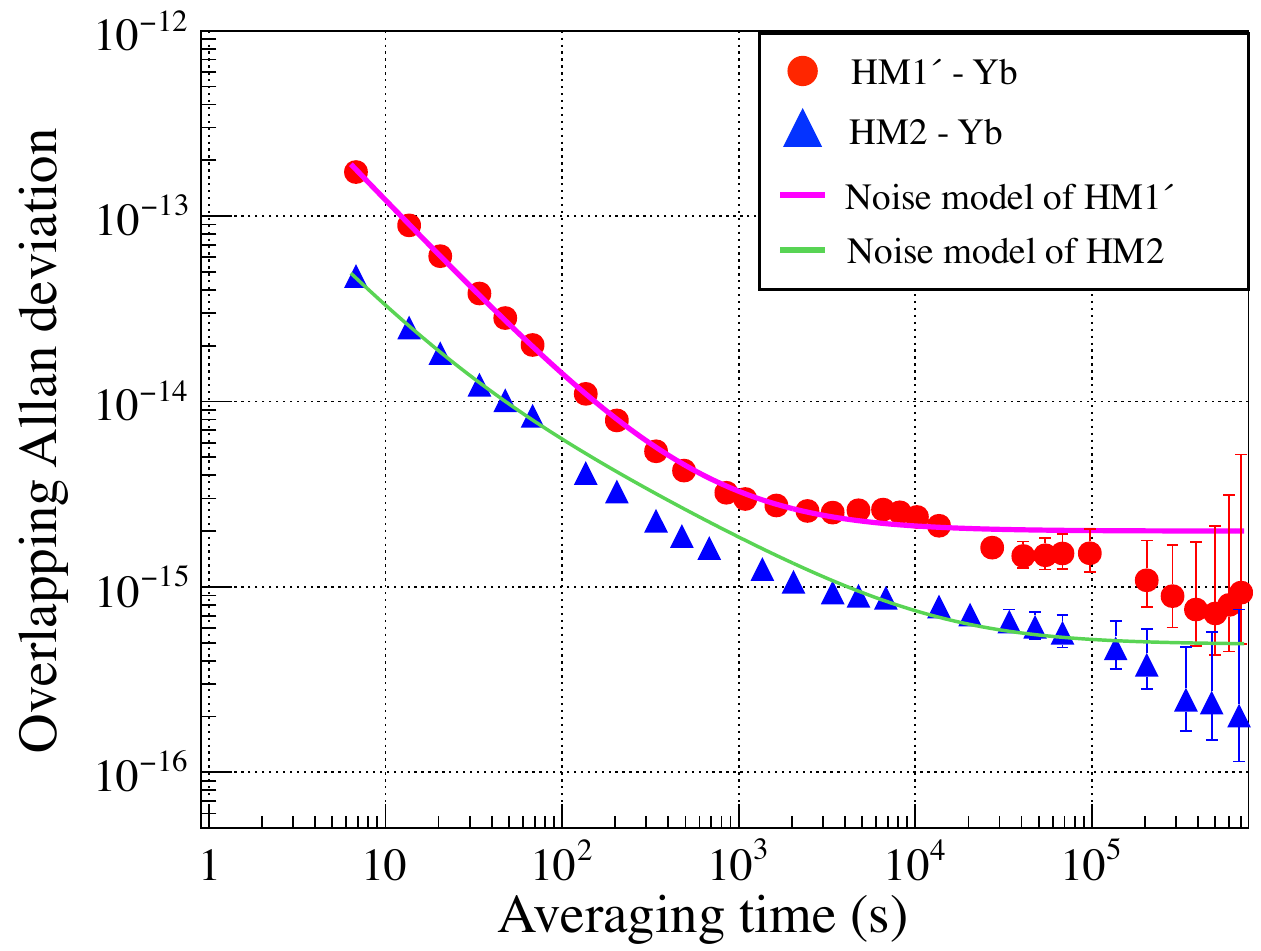}
\caption{Overlapping Allan deviations of HM1$^{\prime}$ (red dot) and HM2 (blue triangle) referenced to NMIJ-Yb1. The red dots are calculated using the data taken from MJD 58754 to MJD 58778 \cite{Kobayashi2020}, and the blue triangles are from MJD 60039 to MJD 60069. The error bar indicates the 95 $\%$ confidence interval. The purple and green lines indicate the noise models of HM1$^{\prime}$ and HM2, respectively. For the generation of UTC(NMIJ)$^{\prime}$ from MJD 58799 to MJD 59029, HM1$^{\prime}$ was employed. The noise model of HM2 shown here is used to simulate a projected performance of UTC(NMIJ)$^{\prime}$ (see Sect.~\ref{discussionsection}).}
\label{maserstability}
\end{figure}

\subsection{Frequency steering scheme}
\label{steeringschemesection}
In this section, we describe a scheme for the frequency steering of UTC(NMIJ)$^{\prime}$ performed in a post-processing analysis, but note that this scheme is applicable to real time generation of UTC(NMIJ)$^{\prime}$. The frequency steering of UTC(NMIJ)$^{\prime}$ is carried out using a Kalman filter algorithm \cite{Yao2018} so that the frequency difference between UTC(NMIJ)$^{\prime}$ and NMIJ-Yb1 is close to zero. The details of the Kalman filter are described in Appendix A. The Kalman filter estimates the fractional frequency offset $y^{\mathrm{e}}_{i}$ and the frequency drift $d^{\mathrm{e}}_{i}$ of HM1$^{\prime}$ against NMIJ-Yb1 at an epoch $i$. For steering UTC(NMIJ)$^{\prime}$ at a next epoch $i+1$, a correction frequency $\Delta y_{i+1}^{\mathrm{Kalman}}$ which is added to the frequency of AOG$^{\prime}$ is determined by 
\begin{equation}
\Delta y_{i+1}^{\mathrm{Kalman}}=-(y^{\mathrm{e}}_{i}+d^{\mathrm{e}}_{i}\Delta t),
\label{kalmancorrection}
\end{equation}
where $\Delta t$ denotes the time interval between epochs. Before the estimation of $y^{\mathrm{e}}_{i}$ and $d^{\mathrm{e}}_{i}$, the Kalman filter firstly predicts the frequency offset $y^{\mathrm{p}}_{i}$ and the drift $d^{\mathrm{p}}_{i}$ based on $y^{\mathrm{e}}_{i-1}$ and $d^{\mathrm{e}}_{i-1}$ estimated at a previous epoch $i-1$. When the measured value $y^{\mathrm{m}}_{i}$ in Eq.~(\ref{hm1ybeq}) is available, $y^{\mathrm{p}}_{i}$ and $d^{\mathrm{p}}_{i}$ are updated to yield better estimates $y^{\mathrm{e}}_{i}$ and $d^{\mathrm{e}}_{i}$ by a weighted average of the predicted and measured values with their weights determined based on the noise characteristics of HM1$'$. During the dead time of NMIJ-Yb1, this update of the predicted values is not performed. 

To determine $\Delta t$ and the weights to update $y^{\mathrm{p}}_{i}$ and $d^{\mathrm{p}}_{i}$, the frequency instability of HM1$^{\prime}$ is evaluated by comparing it with NMIJ-Yb1 before the generation of UTC(NMIJ)$^{\prime}$, which is shown in Figure \ref{maserstability}. To make use of the observed instability in the Kalman filter, we assume a noise model of HM1$^{\prime}$ that approximately characterizes the Allan deviation of HM1$^{\prime}$ with four noise parameters: $1\times10^{-12}/(\tau/\mathrm{s})$ for the white phase modulation (PM), $7\times10^{-14}/\sqrt{(\tau/\mathrm{s})}$ for the white frequency modulation (FM), $2\times10^{-15}$ for the flicker FM, and $4\times10^{-24}\sqrt{(\tau/\mathrm{s})}$ for the random walk FM. Taking account of the fact that the white PM and FM dominate at $\tau\lesssim1\times10^{3}$ s, we choose $\Delta t=1\times10^{3}$ s and calculate the average value of $y^{\mathrm{m}}_{i}$ over $\Delta t$. The determination of the weights to update $y^{\mathrm{p}}_{i}$ and $d^{\mathrm{p}}_{i}$ requires the estimation of the process noise and the measurement noise (see Appendix A). Here a variance $Q^{11}$ characterizing the process noise of the frequency offset is estimated from the flicker FM as $Q^{11}=(2\times10^{-15})^{2}$, while a variance $R_{i}$ describing the measurement noise of the frequency offset is estimated by the white PM and FM, i.e., $R_{i}=(1\times10^{-12}/(\tau_{i}/\mathrm{s}))^{2}+(7\times10^{-14}/\sqrt{(\tau_{i}/\mathrm{s})})^{2}$ with $\tau_{i}$ equal to the uptime of NMIJ-Yb1 for $\Delta t$ at the epoch $i$. The process noise of the frequency drift characterized by a variance $Q^{22}$ is estimated as $Q^{22}=(3\times10^{-24}$ s$^{-1})^{2}$ based on the standard deviation of the monthly drift values of HM1 referenced to TAI for a previous year before the generation of UTC(NMIJ)$^{\prime}$, which is provided by BIPM \cite{CircularT}, and an assumption of a random walk behavior $Q^{22}\propto\tau$. 

Since the frequency of UTC is not always close to that of the SI second, the frequency steering based on a primary or secondary frequency standard can increase the time difference between a steered local time scale and UTC in a long term \cite{Hachisu2018}. For example, during the period of the generation of UTC(NMIJ)$^{\prime}$, the fractional frequency offset of UTC against an ensemble of primary and secondary frequency standards, which is provided by BIPM \cite{CircularT}, is about $5\times10^{-16}$, causing a time difference of $-9$ ns after 200 days. To address this issue, a correction frequency $y^{\mathrm{UTC-Yb}}_{i}$ which approximately corresponds to the fractional frequency difference between UTC and NMIJ-Yb1 is added to $\Delta y_{i}^{\mathrm{Kalman}}$, 
\begin{equation}
\Delta y_{i}^{\mathrm{AOG^{\prime}}}=\Delta y_{i}^{\mathrm{Kalman}}+y^{\mathrm{UTC-Yb}}_{i},
\end{equation}
where $\Delta y_{i}^{\mathrm{AOG^{\prime}}}$ denotes the updated correction frequency applied to AOG$^{\prime}$. Based on discussions of Ref.~\cite{Bauch2012}, we determine $y^{\mathrm{UTC-Yb}}_{i}$ in the following method. The frequency difference between UTC and NMIJ-Yb1 averaged over a previous month is calculated based on the frequency difference between UTC and UTC(NMIJ) provided by BIPM on around 10th of each month \cite{CircularT}. This frequency difference value between UTC and NMIJ-Yb1 is assigned to $y^{\mathrm{UTC-Yb}}_{i}$ in a current epoch $i$. $y^{\mathrm{UTC-Yb}}_{i}$ is updated once a month when a new frequency difference between UTC and UTC(NMIJ) for a previous month is provided by BIPM. 

\section{Results}
\label{resultssection}
\begin{figure*}[t]
\includegraphics[scale=0.48,angle=0]{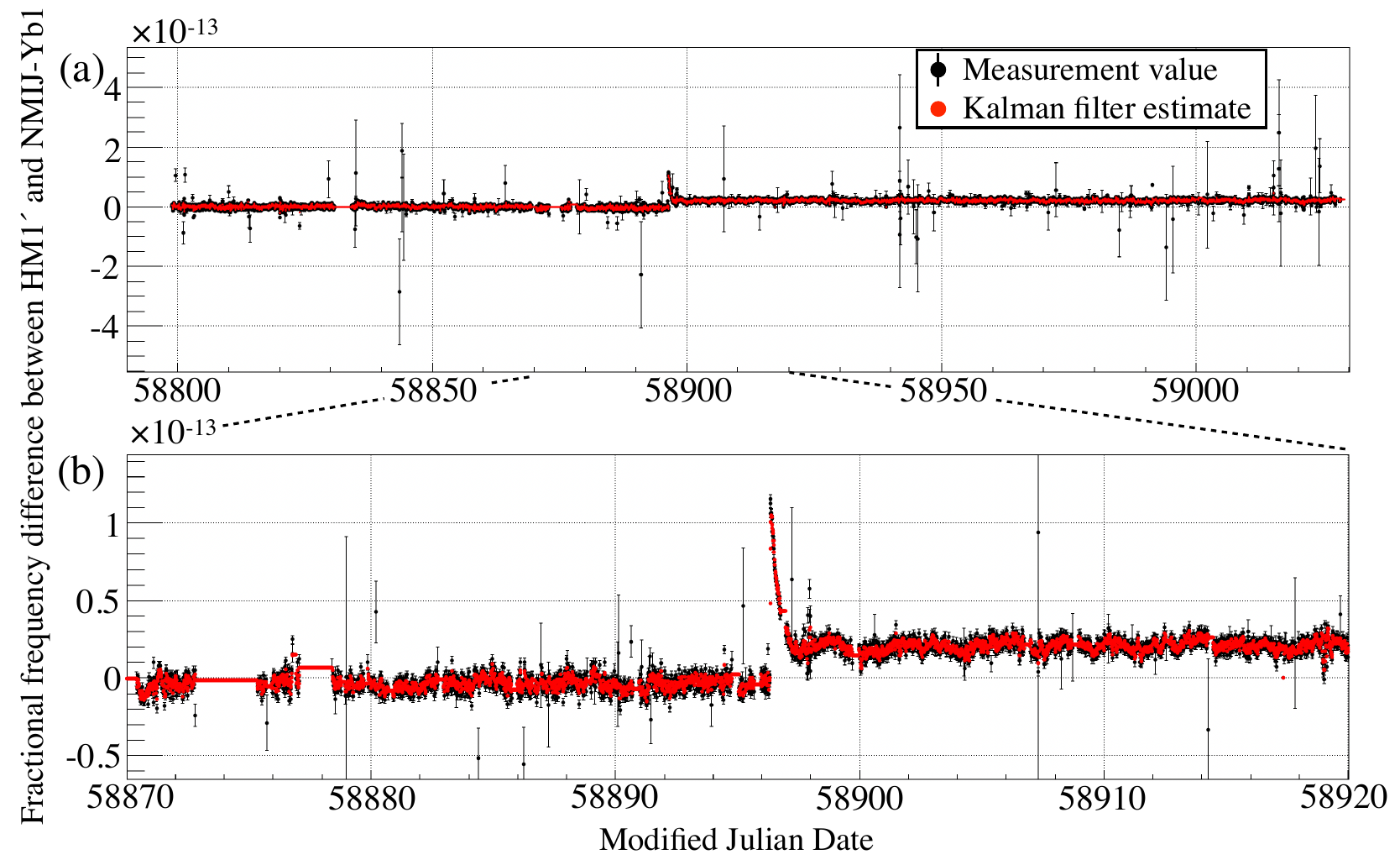}
\caption{Fractional frequency difference between HM1$^{\prime}$ and NMIJ-Yb1 during (a) an entire campaign period of 230 days from MJD 58799 to MJD 59029 and (b) a specific period from MJD 58870 to MJD 58920, which shows an enlarged view of a large frequency fluctuation caused by the maintenance at MJD 58896. The black point shows the measured value averaged over an interval of $\Delta t = 1\times10^{3}$ s and the red point the estimated value by the Kalman filter. The error bar of the measured value indicates the measurement noise $\sqrt{R_{i}}$ used in the Kalman filter estimate, which is calculated by the white noise components of HM1$^{\prime}$.}
\label{uptimedata}
\end{figure*}
\begin{figure}[t]
\includegraphics[scale=0.35,angle=0]{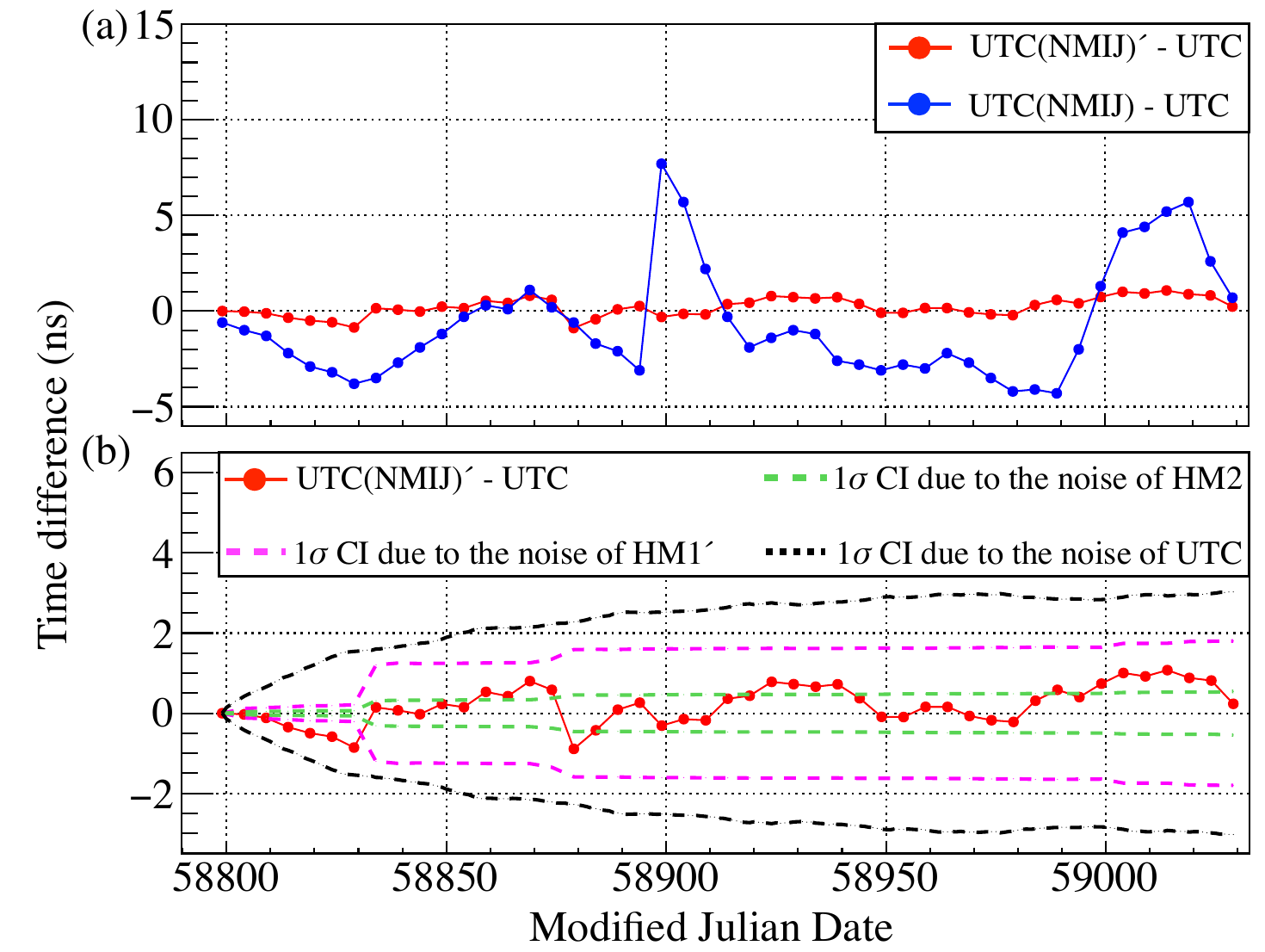}
\caption{(a) Time differences between UTC(NMIJ)$^{\prime}$ (red point) and UTC(NMIJ) (blue point) from UTC. The time offset of UTC(NMIJ)$^{\prime}$ with respect to UTC is initially set to 0 ns at MJD 58799. (b) Enlarged view of the time difference between UTC(NMIJ)$^{\prime}$ and UTC, together with expected $1\sigma$ confidence intervals (CIs) of the time difference due to the noises of HM1$^{\prime}$ (pink dashed line) and HM2 (green dashed line) during the dead time of NMIJ-Yb1, and due to the noise of UTC with the effect of the monthly frequency adjustment (black dashed line).}
\label{timediffresult}
\end{figure}

UTC(NMIJ)$^{\prime}$ was generated during a campaign period of 230 days from Modified Julian Date (MJD) 58799 (12 November 2019) to MJD 59029 (29 June 2020). During this period, NMIJ-Yb1 was operated with an uptime of 81.6 $\%$. Technical details about the robustness of NMIJ-Yb1 are described elsewhere \cite{Kobayashi2020,Kobayashi2019,Hisai2019}. Before the campaign, we operated NMIJ-Yb1 with an uptime of 93.9 $\%$ for 24 days from MJD 58754 to MJD 58778 (i) to evaluate the noise characteristics of HM1$^{\prime}$ and (ii) to obtain the frequency difference between UTC and NMIJ-Yb1 (see Sect.~\ref{steeringschemesection}). The start date of MJD 58799 is set after the first correction frequency $y^{\mathrm{UTC-Yb}}_{i}$ is determined based on the information provided by BIPM. Figure \ref{uptimedata} (a) shows the measured $y^{\mathrm{m}}_{i}$ averaged over $\Delta t$, together with the Kalman filter estimate $y^{\mathrm{e}}_{i}$. At MJD 58896, a large excursion of the phase of UTC(NMIJ) occurred during its maintenance, which was observed by comparing UTC(NMIJ) with another hydrogen maser HM2 (VREMYA VCH-1003M) and a commercial Cs clock. This phase excursion was measured with a time interval counter as $-78.6$ ns referenced to HM2 (see Fig.~\ref{experimentalsetup}), and was corrected in the computation of HM1$^{\prime}$. The frequency of HM1$^{\prime}$ also largely fluctuated at this time, but this fluctuation was well tracked by the Kalman filter thanks to the nearly continuous operation of NMIJ-Yb1, which is shown in Fig.~\ref{uptimedata} (b).

To evaluate the performance of UTC(NMIJ)$^{\prime}$, UTC(NMIJ)$^{\prime}$ was compared with UTC via a link based on global navigation satellite systems (GNSS). The time difference between UTC(NMIJ)$^{\prime}$ and UTC is obtained by the relationship (see Fig. \ref{experimentalsetup})
\begin{eqnarray}
x(\mathrm{UTC(NMIJ)^{\prime}-UTC})&=&x(\mathrm{UTC(NMIJ)^{\prime}-HM1^{\prime}})\nonumber\\
&+&x(\mathrm{HM1^{\prime}-UTC(NMIJ))})\nonumber\\
&+&x(\mathrm{UTC(NMIJ)-UTC}),
\label{taybutc}
\end{eqnarray}
where $x(\mathrm{A-B})$ denotes the time difference between A and B. Since BIPM provides $x(\mathrm{UTC(NMIJ)-UTC})$ at 5-day intervals \cite{CircularT}, we calculate $x(\mathrm{UTC(NMIJ)^{\prime}-HM1^{\prime}})$ and $x(\mathrm{HM1^{\prime}-UTC(NMIJ))})$ at the same intervals by numerically integrating the correction frequencies,
\begin{equation}
x(\mathrm{UTC(NMIJ)^{\prime}-HM1^{\prime}})=\sum_{i=i_{0}}^{i_{\mathrm{f}}} \Delta y_{i}^{\mathrm{AOG^{\prime}}} \Delta t,
\end{equation}
\begin{equation}
x(\mathrm{HM1^{\prime}-UTC(NMIJ)})=-\sum_{i=i_{0}}^{i_{\mathrm{f}}} \Delta y_{i}^{\mathrm{AOG}} \Delta t,
\end{equation}
where times at the initial and final epochs $i_{0}$ and $i_{\mathrm{f}}$ coincide with times at which $x(\mathrm{UTC(NMIJ)-UTC})$ is provided by BIPM. 

Figures \ref{timediffresult} (a) and (b) show the result of $x(\mathrm{UTC(NMIJ)^{\prime}-UTC})$ during the 230-day campaign period. The behavior of our official time scale $x(\mathrm{UTC(NMIJ)-UTC})$ during the same period is also shown in Fig.~\ref{timediffresult} (a). It is worth noting here that NMIJ-Yb1 partially contributed to reduce the time deviation of UTC(NMIJ) from UTC at around MJD 58896 when the large frequency excursion occurred. The time variation of UTC(NMIJ)$^{\prime}$ is significantly improved compared with that of UTC(NMIJ). The root mean square variation of UTC(NMIJ)$^{\prime}$ with respect to UTC is 0.52 ns with a peak-to-peak variation of 2.0 ns. A maximum time deviation of UTC(NMIJ)$^{\prime}$ from UTC is 1.1 ns. 

\begin{figure}[t]
\includegraphics[scale=0.38,angle=0]{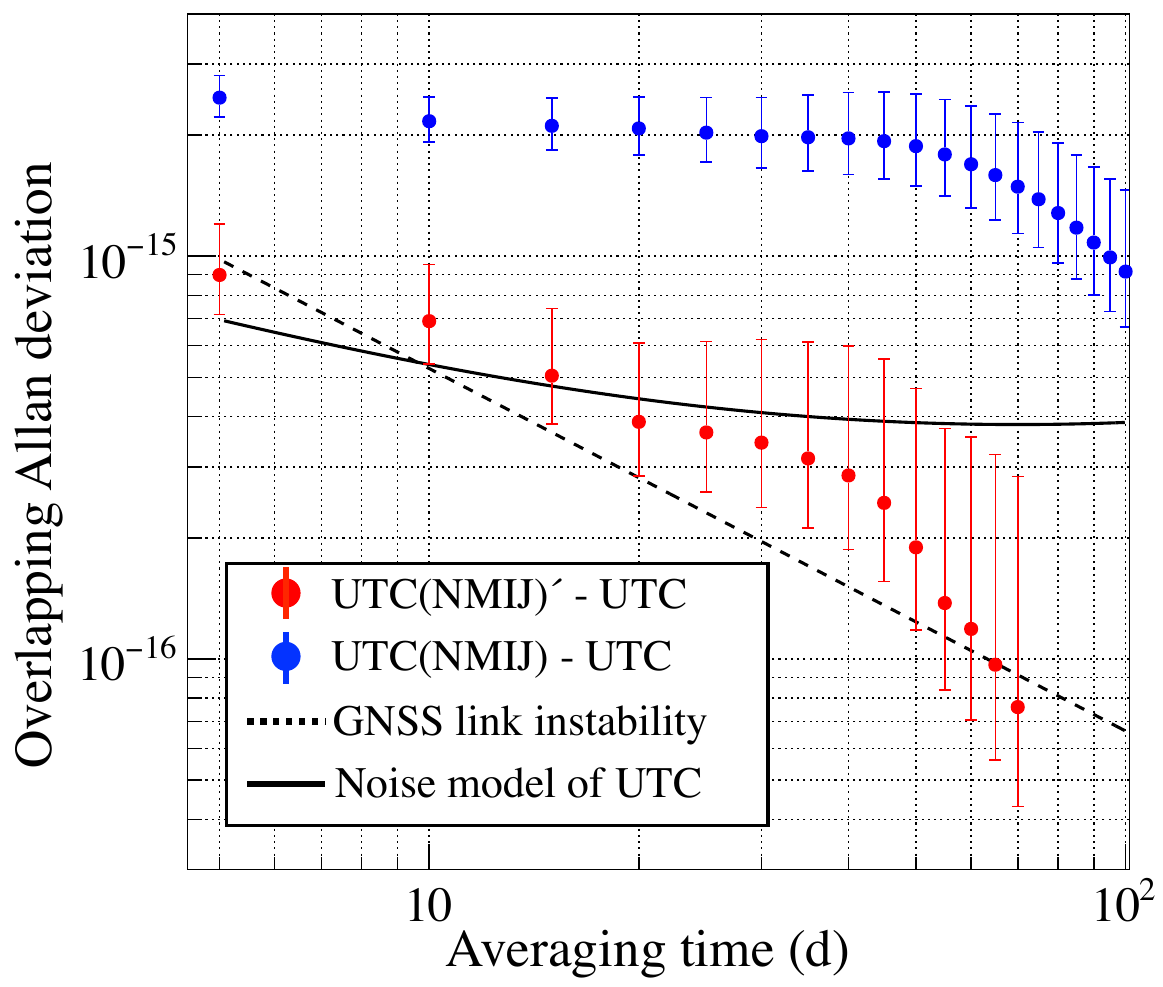}
\caption{Overlapping Allan deviation of UTC(NMIJ)$^{\prime}$ referenced to UTC (red point). A typical frequency stability of UTC(NMIJ) against UTC calculated with the time comparison data from MJD 58004 to MJD 59039 is also shown (blue point). The error bar indicates the 95 $\%$ confidence interval. The black dashed and solid lines indicate the GNSS link instability and the noise model of UTC, respectively, estimated by BIPM \cite{linkcalculation,utcnoise}.}
\label{stabilityresult}
\end{figure}

We also analyzed the frequency stability of the obtained data of $x(\mathrm{UTC(NMIJ)^{\prime}-UTC})$. Figure \ref{stabilityresult} shows the overlapping Allan deviation of UTC(NMIJ)$^{\prime}$ referenced to UTC. For comparison, the GNSS link instability \cite{linkcalculation} and the noise model of UTC \cite{utcnoise} estimated by BIPM are also shown in Fig.~\ref{stabilityresult}. At $\tau\lesssim10$ d, the Allan deviation of $x(\mathrm{UTC(NMIJ)^{\prime}-UTC})$ is mostly dominated by the GNSS link noise. The link noise becomes negligible at $\tau\gtrsim10$ d due to its $1/\tau^{0.9}$ dependence \cite{Panfilo2010}. At $\tau\sim10-40$ d, the Allan deviation reaches a flicker floor of $\sim4\times10^{-16}$. We attribute this floor to the instability of UTC, since the Allan deviation approximately follows the noise model of UTC. Other possibilities are discussed in Sect.~\ref{discussionsection}. At $\tau\gtrsim40$ d, the Allan deviation further decreases due to the monthly frequency adjustment of UTC(NMIJ)$^{\prime}$ by $y^{\mathrm{UTC-Yb}}_{i}$ (see Sect.~\ref{steeringschemesection}), i.e., the frequency of UTC(NMIJ)$^{\prime}$ is locked to that of UTC in the long term.

\section{Discussions and conclusion}
\label{discussionsection}
\begin{figure}[t]
\includegraphics[scale=0.31,angle=0]{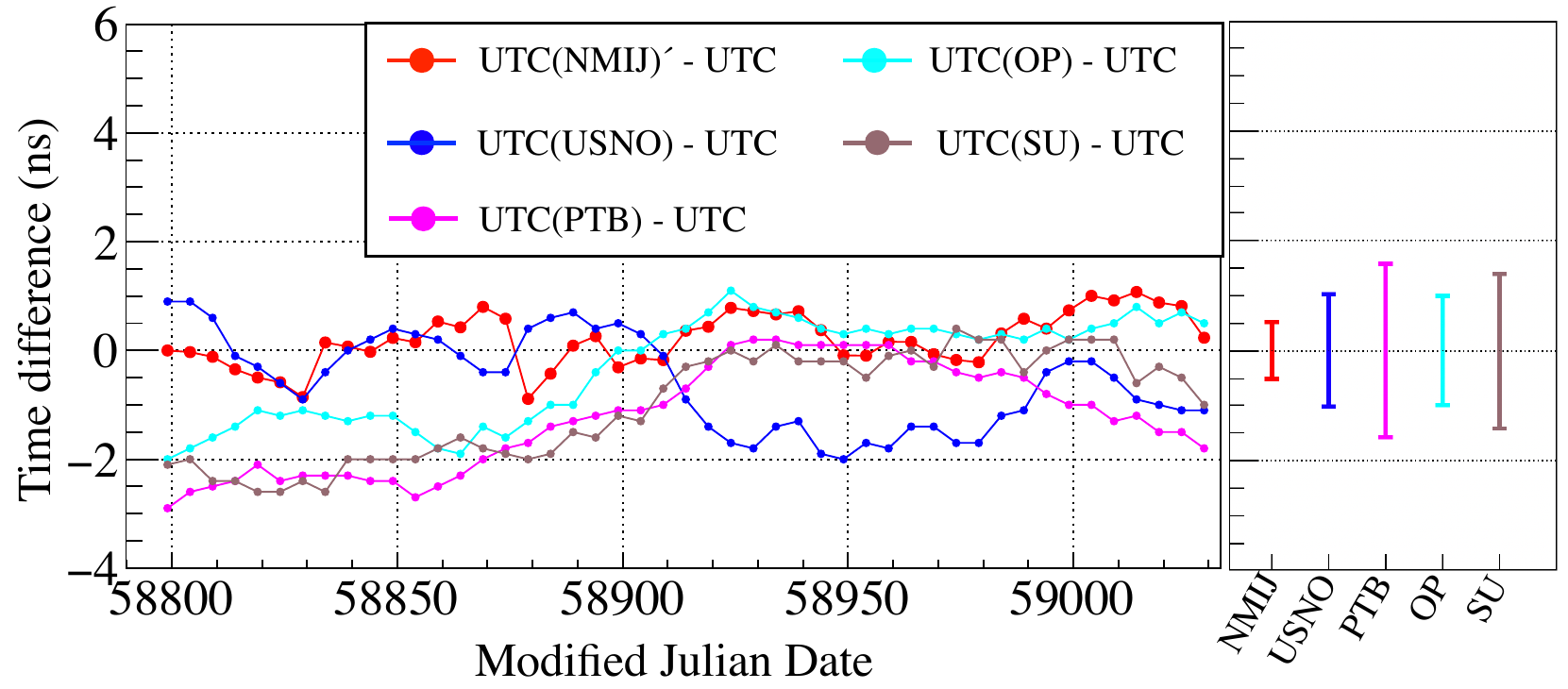}
\caption{Time differences of UTC($k$)s from UTC during the campaign period of the generation of UTC(NMIJ)$^{\prime}$. The error bar in the right figure indicates the root mean square time variation of UTC($k$) with respect to UTC. USNO: United States Navel Observatory, PTB: Physikalisch-Technische Bundesanstalt, OP: Observatoire de Paris, SU: Soviet Union.}
\label{utccomparison}
\end{figure}

The performance of UTC(NMIJ)$^{\prime}$ is compared with those of UTC($k$)s in four institutes based on Cs or Rb microwave fountain clocks \cite{Bauch2012,Domnin2013,Peil2014,Rovera2016} operated during our 230-day campaign period, which are examples of state-of-the-art UTC($k$)s generated in real time for many years. Figure \ref{utccomparison} shows the time differences of these fountain-based UTC($k$)s from UTC \cite{CircularT}. The root mean square time variations of these UTC($k$)s with respect to UTC range between 0.99 ns and 1.6 ns, and the peak-to-peak variations are $\sim3$ ns. Compared with these UTC($k$)s, UTC(NMIJ)$^{\prime}$ exhibits a better performance.

To estimate the effect of the dead time of NMIJ-Yb1 on the performance of UTC(NMIJ)$^{\prime}$, we carried out a Monte Carlo simulation of UTC(NMIJ)$^{\prime}$ with the noise model of HM1$^{\prime}$ (see Sect.~\ref{experimentalsetupsection}) and derived the time difference of UTC(NMIJ)$^{\prime}$ from an ideal (noiseless) reference time scale. The details of the simulation are described in Appendix B. Note that this simulation does not include the instabilities arising from NMIJ-Yb1, the comb, the GNSS link, and UTC. A $1\sigma$ confidence interval of the time difference obtained by the simulation is shown in Fig.~\ref{timediffresult} (b). During the first 30 days from MJD 58799 to MJD 58829 with a high uptime of 88.3 $\%$, the absolute value of the time difference is only $\leq0.2$ ns. Over the entire campaign period, however, the 1$\sigma$ confidence interval rapidly increases to a range between $\pm$1.2 ns at MJD 58834 and $\pm1.6$ ns at MJD 58879, and finally reaches $\pm1.8$ ns at MJD 59029. The rapid changes are caused by the flicker noise of HM1$^{\prime}$ and the prediction error of the Kalman filter during two large dead times of $\sim3$ days from MJD 58830 and $\sim2$ days from MJD 58873.

The real data $x(\mathrm{UTC(NMIJ)^{\prime}-UTC})$ shown in Fig.~\ref{timediffresult} (b) mostly stay within the 1$\sigma$ confidence interval, but deviate outside the interval during the first 30 days, implying that the dead time of NMIJ-Yb1 is not a dominant source that limits the stability of UTC(NMIJ)$^{\prime}$ during this 30-day period. Since the frequency stability of the real data is expected to be limited by that of UTC at $\tau\sim10-40$ d (see Sect.~\ref{resultssection}), it is likely that this deviation is mainly due to the instability of UTC. This is also confirmed by comparing the real data with a $1\sigma$ confidence interval calculated from the noise model of UTC (see Fig.~\ref{timediffresult} (b)). Other possibilities include the noise contributions from NMIJ-Yb1 and the comb-based measurement. The noise of NMIJ-Yb1 is assumed to be negligibly small, since the observed frequency stability of NMIJ-Yb1 reaches $\sim1\times10^{-16}$ at $\tau\sim5\times10^{3}$ s with a $1/\sqrt{\tau}$ slope \cite{Kobayashi2020}. The noise of the comb-based measurement may have a small contribution if the observed flicker noise of $2.2\times10^{-16}$ at $\tau\sim(1-5)\times10^{4}$ s (see Sect.~\ref{experimentalsetupsection}) is still dominant at $\tau\sim30$ d. 

With a better flywheel oscillator, unprecedented performance of UTC(NMIJ)$^{\prime}$ is expected thanks to the high uptime of NMIJ-Yb1. To demonstrate this, we carried out another Monte Carlo simulation using a noise model of HM2 which exhibits a better frequency stability: $3\times10^{-13}/(\tau/\mathrm{s})$ for the white PM, $6\times10^{-14}/\sqrt{(\tau/\mathrm{s})}$ for the white FM, $5\times10^{-16}$ for the flicker FM, and $2\times10^{-27}\sqrt{(\tau/\mathrm{s})}$ for the random walk FM (see Fig.~\ref{maserstability}). The frequency stability of HM2 was evaluated against NMIJ-Yb1 after the campaign by establishing a direct link between NMIJ-Yb1 and HM2. We note that HM2 was not employed as a flywheel oscillator during the campaign, since the frequency stability of the comparison between HM2 and NMIJ-Yb1 was heavily degraded by measurement noise of a time interval counter (see Fig.~\ref{experimentalsetup}). With the same uptime of NMIJ-Yb1 (81.6 $\%$ for 230 days), the absolute value of the time difference obtained with the model of HM2 is only $\leq0.06$ ns during the first 30 days, and $\leq0.54$ ns over the entire 230 days, as shown in Fig.~\ref{timediffresult} (b). The performance during the first 30 days is comparable to a reported time error of $48\pm94$ ps against an ideal reference for 34 days achieved by an all-optical time scale based on a cryogenic silicon cavity \cite{Milner2019}.

When the instability arising from the noise of HM2 during the dead time of NMIJ-Yb1 reaches this level, the other noise sources would limit the performance of UTC(NMIJ)$^{\prime}$. The dominant sources are expected to be the GNSS link and UTC as long as UTC(NMIJ)$^{\prime}$ is compared with UTC in the current infrastructure. Considering the stability of UTC(NMIJ)$^{\prime}$ itself, it may be limited by the noises of NMIJ-Yb1 and the comb-based measurement. We have already reduced some of the noises in more recent investigations \cite{Kobayashi2022,Wada2022} and plan to achieve the expected stability level estimated by the simulation in the future. 

In conclusion, we have generated UTC(NMIJ)$^{\prime}$ for 230 days by steering the frequency of a single hydrogen maser to that of NMIJ-Yb1 running nearly continuously. UTC(NMIJ)$^{\prime}$ exhibits a lower time variation with respect to UTC compared with those of other UTC($k$)s based on Cs or Rb fountain clocks. We have demonstrated that a use of an optical clock with a high uptime enhances the stability of a time scale. This work constitutes an essential step towards a redefinition of the SI second.

\section*{Acknowledgements}
We dedicate this paper to the memory of T. Suzuyama who made an essential contribution to this paper but passed away on 4 July 2020. We are indebted to A. Iwasa and Y. Fujii for the support to maintain UTC(NMIJ). We thank the BIPM time department for making the data available in the website. This work was supported by Japan Society for the Promotion of Science (JSPS) KAKENHI Grant Number 17H01151, 17K14367, 22H01241, JST-Mirai Program Grant Number JPMJMI18A1, and JST Moonshot R$\&$D Program Grant Number JPMJMS2268, Japan.

\section*{Appendix A: Kalman filter algorithm}
We employ a Kalman filter algorithm similar to that described in Ref.~\cite{Yao2018}. The Kalman filter estimates the frequency offset $y_{i}^{\mathrm{e}}$ and the frequency drift $d_{i}^{\mathrm{e}}$ of the hydrogen maser against NMIJ-Yb1 at a time epoch $i$ with a time interval $\Delta t$ to yield the correction frequency applied to AOG$^{\prime}$ at a next epoch $i+1$ (see Eq.~(\ref{kalmancorrection})). Before $y_{i}^{\mathrm{e}}$ and $d_{i}^{\mathrm{e}}$ are determined, the Kalman filter predicts the frequency offset $y_{i}^{\mathrm{p}}$ and the drift $d_{i}^{\mathrm{p}}$ based on $y_{i-1}^{\mathrm{e}}$ and $d_{i-1}^{\mathrm{e}}$ estimated at a previous epoch $i-1$ by
\begin{equation}
\renewcommand{\arraystretch}{1.5}
\left(
\begin{array}{c}
y_{i}^{\mathrm{p}} \\
d_{i}^{\mathrm{p}} 
\end{array}
\right)
=
\left(
\begin{array}{cc}
1 & \Delta t \\
0 & 1 
\end{array}
\right)
\left(
\begin{array}{c}
y_{i-1}^{\mathrm{e}} \\
d_{i-1}^{\mathrm{e}}. 
\end{array}
\right).
\label{linearmodel}
\end{equation}
During the dead time of NMIJ-Yb1, the predicted values are simply used for $y_{i}^{\mathrm{e}}$ and $d_{i}^{\mathrm{e}}$, i.e.,
\begin{equation}
\renewcommand{\arraystretch}{1.5}
\left(
\begin{array}{c}
y_{i}^{\mathrm{e}} \\
d_{i}^{\mathrm{e}} 
\end{array}
\right)
=
\left(
\begin{array}{c}
y_{i}^{\mathrm{p}} \\
d_{i}^{\mathrm{p}} 
\end{array}
\right).
\end{equation}
When the measured frequency offset $y_{i}^{\mathrm{m}}$ by NMIJ-Yb1 is available, the predicted values are updated by a weighted average of the predicted and measured values as follows,
\begin{equation}
\renewcommand{\arraystretch}{1.5}
\left(
\begin{array}{c}
y_{i}^{\mathrm{e}} \\
d_{i}^{\mathrm{e}} 
\end{array}
\right)
=
\left(
\begin{array}{c}
y_{i}^{\mathrm{p}} \\
d_{i}^{\mathrm{p}} 
\end{array}
\right)
+
K_{i}(y_{i}^{\mathrm{m}}-y_{i}^{\mathrm{p}}),
\end{equation}
where $K_{i}$ is the Kalman gain matrix which determines the weights of the predicted and measured values. In the present model with the frequency offset and the drift as described in Eq.~(\ref{linearmodel}), the Kalman gain matrix is given by
\begin{equation}
\renewcommand{\arraystretch}{1.5}
K_{i}=
\left(
\begin{array}{c}
K^{y}_{i} \\
K^{d}_{i} 
\end{array}
\right)
=
\left(
\begin{array}{c}
\frac{p^{11}_{i} + \Delta t (p^{21}_{i}+p^{12}_{i} + \Delta t p^{22}_{i})+Q^{11}}{p^{11}_{i} + \Delta t (p^{21}_{i} +p^{12}_{i} + \Delta t p^{22}_{i}) + Q^{11} + R_{i}} \\
\frac{p^{21}_{i}+\Delta t p^{22}_{i}}{p^{11}_{i} + \Delta t(p^{21}_{i} +p^{12}_{i} + \Delta t p^{22}_{i}) + Q^{11} + R_{i}}
\end{array}
\right),
\label{kalmangainelements}
\end{equation}
where $Q^{11}$ and $R_{i}$ are variances characterizing the process noise and the measurement noise of the frequency offset, respectively. $p^{11}_{i}$, $p^{12}_{i}$, $p^{21}_{i}$, and $p^{22}_{i}$ are elements of a covariance matrix $P_{i}$ of the estimate, defined as 
\begin{equation}
\renewcommand{\arraystretch}{1.5}
P_{i}
=
\left(
\begin{array}{cc}
p_{i}^{11} & p_{i}^{12} \\
p_{i}^{21} & p_{i}^{22}
\end{array}
\right),
\end{equation}
and are obtained from $P_{i-1}$ and $K_{i-1}$ at the previous epoch $i-1$ by
\begin{eqnarray}
p_{i}^{11} &=&  (1 - K^{y}_{i-1})[p_{i-1}^{11} + \Delta t (p_{i-1}^{21}+p_{i-1}^{12} + \Delta t p_{i-1}^{22}) + Q^{11}], \nonumber\\
p_{i}^{12} &=& (1 - K_{i-1}^{y})(p_{i-1}^{12} + \Delta t p_{i-1}^{22}), \nonumber\\
p_{i}^{21} &=& p_{i-1}^{21} + \Delta t p_{i-1}^{22} \nonumber\\
&&- K_{i-1}^{d} [p^{11}_{i-1} + \Delta t(p^{21}_{i-1} +p^{12}_{i-1} + \Delta t p^{22}_{i-1}) + Q^{11}],\nonumber\\
p_{i}^{22} &=& p^{22}_{i-1} - K_{i-1}^{d}(p^{12}_{i-1} + \Delta t p^{22}_{i-1}) + Q^{22},
\end{eqnarray}
where $Q^{22}$ is a variance characterizing the process noise of the drift.
The execution of the Kalman filter algorithm requires the determination of $Q^{11}$, $Q^{22}$, and $R_{i}$ by user. We estimate these variances based on the measured noise characteristics of the hydrogen maser (see Sect.~\ref{steeringschemesection}).

\section*{Appendix B: Monte Carlo simulation}
In the Monte Carlo simulation, we generate time series data of the frequency with a power spectral density which is converted from the Allan deviation characterized by the noise model of HM1$^{\prime}$ or HM2 \cite{Handbook}. The frequency data are then gapped according to the dead time of NMIJ-Yb1. With the same Kalman filter algorithm, the frequency of HM1$^{\prime}$ or HM2 is estimated from the gapped data. By integrating the frequency difference between the frequency estimated by the Kalman filter and the true frequency obtained from the data without the gap, the time difference between the simulated UTC(NMIJ)$^{\prime}$ and an ideal time scale is obtained. We repeat this procedure 200 times and calculate the root mean square of the time differences, which is the $1\sigma$ confidence interval of the time difference shown in Fig.~\ref{timediffresult} (b). 

For calculation of the $1\sigma$ confidence interval due to the noise of UTC, we generate another Monte Carlo simulation data set of the frequency based on the noise model of UTC \cite{utcnoise}. To include the effect of the monthly frequency adjustment of UTC(NMIJ)$^{\prime}$ by $y^{\mathrm{UTC-Yb}}_{i}$ (see Sect.~\ref{steeringschemesection}), a correction frequency is similarly calculated approximately once a month by averaging simulated frequencies over a previous month, and is added to a simulated frequency at a current time epoch. The resulting frequencies are integrated to obtain the time difference between UTC and an ideal time scale.


\begin{thebibliography}{1}
\bibitem{Takano2016} T. Takano, M. Takamoto, I. Ushijima, N. Ohmae, T. Akatsuka, A. Yamaguchi, Y. Kuroishi, H. Munekane, B. Miyahara, and H. Katori, Nat. Photonics $\bf{10}$, 662 (2016).
\bibitem{Grotti2018} J. Grotti $et$. $al$., Nat. Phys. $\bf{14}$, 437 (2018).
\bibitem{Hees2016} A. Hees, J. Gu\'ena, M. Abgrall, S. Bize, and P. Wolf, Phys. Rev. Lett. $\textbf{117}$, 061301 (2016).
\bibitem{Wcilso2018} P. Wsci\l o $et$ $al.,$ Sci. Adv. $\textbf{4}$, eaau4869 (2018).
\bibitem{Ashby2018} N. Ashby. T. E. Parker, and B. R. Patla, Nat. Phys. $\bf{14}$, 822-826 (2018). 
\bibitem{Kennedy2020} C. J. Kennedy, E. Oelker, J. M. Robinson, T. Bothwell, D. Kedar, W. R. Milner, G. E. Marit, A. Derevianko, and J. Ye, Phys. Rev. Lett. $\textbf{125}$, 201302 (2020).
\bibitem{Takamoto2020} M. Takamoto, I. Ushijima, N. Ohmae, T. Yahagi, K. Kokado, H. Shinkai, and H. Katori, Nat. Photon. $\textbf{14}$, 411-415 (2020).
\bibitem{Lange2021} R. Lange, N. Huntemann, J. M. Rahm, C. Sanner, H. Shao, B. Lipphardt, Chr. Tamm, S. Weyers, and E. Peik, Phys. Rev. Lett. $\textbf{126}$, 011102 (2021).
\bibitem{Kobayashi2022} T. Kobayashi, A. Takamizawa, D. Akamatsu, A. Kawasaki, A. Nishiyama, K. Hosaka, Y. Hisai, M. Wada, H. Inaba, T. Tanabe, and M. Yasuda, Phys. Rev. Lett. $\textbf{129}$, 241301 (2022).
\bibitem{Filzinger2023} M. Filzinger, S. D\"orscher, R. Lange, J. Klose, M. Steinel, E. Benkler, E. Peik, C. Lisdat, and N. Hunteman, Phys. Rev. Lett. $\textbf{130}$, 253001 (2023).
\bibitem{Sherrill2023} A. Sherrill, A. O. Parsons, C. F. A. Baynham, W. Bowden, E. A. Curtis, R. Hendricks, I. R. Hill, R. Hobson, H. S. Margolis, B. I. Robertson, M. Schioppo, K. Szymaniec, A. Tofful, J. Tunesi, R. M. Godun, and X. Calmet, New J. Phys. $\bf{25}$, 093012 (2023).
\bibitem{Panfilo2019} G. Panfilo and F. Arias, Metrologia $\textbf{56}$, 042001 (2019).
\bibitem{CircularT} https://www.bipm.org/en/time-ftp.
\bibitem{Bauch2012} A. Bauch, S. Weyers, D. Piester, E. Staliuniene, and W. Yang, Metrologia $\textbf{49}$, 180 (2012).
\bibitem{Domnin2013} Y. S. Domnin, V. N. Baryshev, A. I. Boyko, G. A. Elkin. A. V. Novoselov, L. N. Kopylov, and D. S. Kupalov, Meas. Tech. $\bf{55}$, 1155-1162 (2013).
\bibitem{Peil2014} S. Peil, J. L. Hanssen, T. B. Swanson, J. Tayler, and C. R. Ekstrom, Metrologia $\textbf{51}$, 263-269 (2014).
\bibitem{Rovera2016} G. D. Rovera, S. Bize, B. Chupin, J. Gu\'ena, P. Laurent, P. Rosenbusch, P. Uhrich, and M. Abgrall, Metrologia $\textbf{53}$, S81 (2016).
\bibitem{Riehle2018} F. Riehle, P. Gill, F. Arias, and L. Robertsson, Metrologia $\bf{55}$, 188 (2018).
\bibitem{Hong2016} F.-L. Hong, Meas. Sci. Technol. $\bf{28}$, 012002 (2016).
\bibitem{Dimarcq2023} N. Dimarcq $et$ $al.,$ Metrologia $\bf{61}$, 012001 (2024). 
\bibitem{Bloom2014} B. J. Bloom, T. L. Nicholson, J. R. Williams, S. L. Campbell, M. Bishof, X. Zhang, W. Zhang, S. L. Bromley, and J. Ye, Nature $\bf{506}$, 71 (2014).
\bibitem{Ushijima2015} I. Ushijima, M. Takamoto, M. Das, T. Ohkubo, and H. Katori, Nat. Photonics $\bf{9}$, 185 (2015).
\bibitem{Huntemann2016} N. Huntemann, C. Sanner, B. Lipphardt, C. Tamm, and E. Peik, Phys. Rev. Lett. $\text{116}$, 063001 (2016).
\bibitem{McGrew2018} W. F. McGrew, X. Zhang, R. J. Fasano, S. A. Sch\"affer, K. Beloy, D. Nicolodi, R. C. Brown, N. Hinkley, G. Milani, M. Schioppo, T. H. Yoon, and A. D. Ludlow, Nature $\bf{564}$, 87 (2018).
\bibitem{Brewer2019} S. M. Brewer, J.-S. Chen, A. M. Hankin, E. R. Clements, C. W. Chou, D. J. Wineland, D. B. Hume, and D. R. Leibrandt, Phys. Rev. Lett. $\textbf{123}$, 033201 (2019).
\bibitem{Hung2022} Y. Hung, B. Zhang, M. Zeng, Y. Hao, Z. Ma, H. Zhang, H. Guan, Z. Chen, M. Wang, and K. Gao, Phys. Rev. Applied $\bf{17}$, 034041 (2022).
\bibitem{Zhang2022} A. Zhang, Z. Xiong, X. Chen, Y. Jiang, J. Wang, C. Tian, Q. Zhu, B. Wang, D. Xiong, L. He, L. Ma, and B. Lyu, Metrologia $\bf{59}$, 065009 (2022).
\bibitem{Zhiqiang2023} Z. Zhiqiang, K. J. Arnold, R. Kaewuam, M. D. Barrett, Sci. Adv. $\bf{9}$, eadg1971 (2023).
\bibitem{Grebing2016} C. Grebing, A. Al-Masoudi, S. D\"orsher, S. H\"afner, V. Gerginov, S. Weyers, B. Lipphardt, F. Riehle, U. Sterr, and C. Lisdat, Optica $\bf{3}$, 563 (2016).
\bibitem{Hachisu2018} H. Hachisu, F. Nakagawa, Y. Hanado, T. Ido, Sci. Rep. $\bf{8}$, 4243 (2018).
\bibitem{Yao2019} J. Yao $et$. $al$., Phys. Rev. Applied $\bf{12}$ 044069 (2019).
\bibitem{Formichella2021} V. Formichella, L. Galleani, G. Signorille, and I. Sesia, Metrologia $\bf{59}$, 015002 (2022).
\bibitem{Zhu2022} L. Zhu, Y. Lin, Y. Wang, Z. Jia, Q. Wang, Y. Li, T. Yang, and Z. Fang, Metrologia $\bf{59}$, 055007 (2022).
\bibitem{Lee2023} H. S. Lee, T. Y. Kwon, S. E. Park, Y. K. Lee, S.-h. Yang, D.-H. Yu, M.-S. Heo, H. Kim, C. Y. Park, and W.-K. Lee, Metrologia (2023).
\bibitem{Xu2023} Q. Xu, S. Wang, X. Lu, F. Guo, J. Xia, Y. Wang, H. Chang, AIP Advances $\bf{13}$, 115316 (2023).
\bibitem{Milner2019} W. R. Milner, J. M. Robinson, C. J. Kennedy, T. Bothwell, D. Kedar, D. G. Matei, T. Legero, U. Sterr, F. Riehle, H. Leopardi, T. M. Fortier, J. A. Sherman, J. Levine, J. Yao, J. Ye, and E. Melker, Phys. Rev. Lett. $\bf{123}$, 173201 (2019).
\bibitem{Hill2016} I. R. Hill, R. Hobson, W. Bowden, E. M. Bridge, S. Donnellan, E. A. Curtis, and P. Gill, J. Phys. Conf. Ser. $\bf{723}$, 012019 (2016).
\bibitem{Lodewyck2016} J. Lodewyck, S. Bilicki, E. Bookjans, J.-L. Robyr, C. Shi, G. Vallet, R. L. Targat, D. Nicolodi, Y. Coq , J. Gu\'ena, M. Abgrall, P. Rosenbusch, and S. Bize, Metrologia $\bf{53}$, 1123 (2016).
\bibitem{Baynham2017} C. F. Baynham, R. M. Godun, J. M. Jones, S. A. King, P. B. R. Nisbet-Jones, F. Baynes, A. Rolland, P. E. G. Baird, K. Bongs, P. Gill, and H. S. Margolis, J. Mod. Opt. $\bf{65}$, 585 (2017).
\bibitem{Stuhler2021} J. Stuhler $et$ $al.,$ Measurement: Sensors $\bf{18}$, 100264 (2021).
\bibitem{Zeng2023} M. Zeng, Y. Huang, B. Zhang, Y. Hao, Z. Ma, R. Hu, H. Zhang, Z. Chen, M. Wang, H. Guan, and K. Gao, Phys. Rev. Applied $\bf{19}$, 064004 (2023).
\bibitem{Kobayashi2020} T. Kobayashi, D. Akamatsu, K. Hosaka, Y. Hisai, M. Wada, H. Inaba, T. Suzuyama, F.-L. Hong, and M. Yasuda, Metrologia $\textbf{57}$, 065021 (2020).
\bibitem{Kobayashi2018} T. Kobayashi, D. Akamatsu, Y. Hisai, T. Tanabe, H. Inaba, T. Suzuyama, F.-L. Hong, K. Hosaka, and M. Yasuda, IEEE Trans. Ultrason., Ferroelect., Freq. Control, $\bf{65}$, 2449 (2018).
\bibitem{Inaba2006} H. Inaba, Y. Daimon, F.-L. Hong, A. Onae, K. Minoshima, T. R. Schibli, H. Matsumoto, M. Hirano, T. Okuno, M. Onishi, and M. Nakazawa, Opt. Express $\bf{14}$, 5223 (2006).
\bibitem{Akamatsu2014} D. Akamatsu, H. Inaba, K. Hosaka, M. Yasuda, A. Onae, T. Suzuyama, M. Amemiya, and F.-L. Hong, Appl. Phys. Express $\bf{7}$, 012401 (2013).
\bibitem{Hisai2021} Y. Hisai, D. Akamatsu, T. Kobayashi, K. Hosaka, H. Inaba, F.-L. Hong, and M. Yasuda, Metrologia $\bf{58}$, 015008 (2021).
\bibitem{stabilitynote} The frequency stability observed in the comparison between the Yb and Sr optical lattice clocks is $\sigma_{\mathrm{Yb/Sr}}=1.0\times10^{-14}/\sqrt{(\tau/\mathrm{s})}$. The frequency stability of the Yb optical lattice clock is derived by $\sigma_{\mathrm{Yb/Sr}}/\sqrt{2}$ assuming that the both clocks have the same frequency stabilities.
\bibitem{Yao2018} J. Yao, T. E. Parker, N. Ashby, and J. Levine, IEEE Trans. Ultrason., Ferroelect., Freq. Control, $\textbf{65}$, 127 (2018).
\bibitem{Hisai2019} Y. Hisai, D. Akamatsu, T. Kobayashi, S. Okubo, H. Inaba, K. Hosaka, M. Yasuda, and F.-L. Hong, Opt. Express $\bf{27}$, 6404 (2019).
\bibitem{Kobayashi2019} T. Kobayashi, D. Akamatsu, K. Hosaka, and M. Yasuda, Rev. Sci. Instrum. $\bf{90}$, 103002 (2019).
\bibitem{linkcalculation} The GNSS link instability is derived from a recommended formula to calculate the satellite link uncertainty $u_{\mathrm{link}}=(\sqrt{2}u_{\mathrm{A}})/[(86400\times5\,\,\mathrm{s})(\frac{(\tau/\mathrm{d})}{5})^{0.9}]$ with the statistical uncertainty $u_{\mathrm{A}}=0.3$ ns \cite{CircularT,Panfilo2010}. 
\bibitem{utcnoise} The noise model of UTC referenced to the SI second is as follows: $1.4\times10^{-15}/\sqrt{(\tau/\mathrm{d})}$ for the white FM, $3\times10^{-16}$ for the flicker FM, and $2\times10^{-17}\sqrt{(\tau/\mathrm{d})}$ for the random walk FM \cite{CircularT}. While BIPM provides this model for the Echelle Atomique Libre (EAL) which corresponds to a flywheel oscillator for TAI and UTC, the frequency offset between EAL and UTC was constant during the campaign period (i.e., the frequency steering of UTC was not performed), and thus the frequency stability of EAL is equivalent to that of UTC.
\bibitem{Panfilo2010} G. Panfilo and T. E. Parker, Metrologia $\bf{47}$, 552 (2010).
\bibitem{Wada2022} M. Wada and H. Inaba, Metrologia $\bf{59}$, 065005 (2022).
\bibitem{Handbook} W. Riley, $Handbook$ $of$ $Frequency$ $Stability$ $Analysis$ (NIST Special Publication, 2008), Vol. 1065.
\end{thebibliography}
\end{document}